\documentclass[lettersize,journal]{IEEEtran}
\usepackage{amsmath,amsfonts}
\usepackage{algorithmic}
\usepackage{algorithm}
\usepackage{array}
\usepackage[caption=false,font=normalsize,labelfont=sf,textfont=sf]{subfig}
\usepackage{textcomp}
\usepackage{stfloats}
\usepackage{url}
\usepackage{verbatim}
\usepackage{graphicx}
\usepackage{cite}
\usepackage{tabu}
\usepackage{pifont}
\usepackage{textcomp}
\usepackage{ntheorem}
\usepackage{tabularx}
\usepackage{booktabs}  
\usepackage{threeparttable}
\usepackage{multicol}  
\usepackage{multirow} 
\usepackage[numbers,sort&compress]{natbib}
\usepackage{epstopdf}
\usepackage[table]{xcolor}
\usepackage{url}
\usepackage{amssymb}
\usepackage{tikz,xcolor}
\definecolor{lime}{HTML}{A6CE39}
\DeclareRobustCommand{\orcidicon}{
	\begin{tikzpicture}
		\draw[lime, fill=lime] (0,0)
		circle[radius=0.16]
		node[white]{{\fontfamily{qag}\selectfont \tiny \.{I}D}}; 
	\end{tikzpicture}
	\hspace{-2mm}
}
\foreach \x in {A, ..., Z}{%
	\expandafter\xdef\csname orcid\x\endcsname{\noexpand\href{https://orcid.org/\csname orcidauthor\x\endcsname}{\noexpand\orcidicon}}
}

\def\BibTeX{{\rm B\kern-.05em{\sc i\kern-.025em b}\kern-.08em
    T\kern-.1667em\lower.7ex\hbox{E}\kern-.125emX}}

\usepackage[colorlinks,
linkcolor=cyan,
anchorcolor=cyan,
citecolor=cyan]{hyperref}

\begin{document}
\title{Flexible Alignment Super-Resolution Network for Multi-Contrast MRI}

\author{
	Yiming Liu\hspace{-1.5mm}\orcidB{}, \IEEEmembership{Member, IEEE}, Mengxi Zhang\hspace{-1.5mm}\orcidA{}, Bo Jiang, Bo Hou,\\ Dan Liu, Jie Chen, \IEEEmembership{Member, IEEE}, Heqing Lian\hspace{-1.5mm}\orcidC{}, \IEEEmembership{Member, IEEE}
	\thanks{
		This work was supported by the China National Funds for Distinguished Young Scientists under Grant 81601485. (\textit{Yiming Liu and Mengxi Zhang contributed equally to this work. Corresponding authors: Jie Chen; Heqing Lian.})
	}
	\thanks{Yiming Liu and Heqing Lian are with the Xiao Ying AI Lab, Beijing, China. ( liuyiming@xiaoyingai.com, lianheqing@xiaoyingai.com).}
	\thanks{Mengxi Zhang is with the School of Electrical and Information Engineering, Tianjin University, Tianjin, China. (mengxizhang@tju.edu.cn).}
	\thanks{Bo Jiang, Dan Liu and Bo Hou are with peking union medical college hospital, Beijing, China. (jbpumch@163.com, liud2104@163.com, houbo97@pumch.cn)}
 \thanks{Jie Chen is with the School of Electronics Engineering and Computer Science, Peking University, China. (jiechen2019@pku.edu.cn)}
}
\maketitle

\begin{abstract}
Magnetic resonance imaging plays an essential role in clinical diagnosis by acquiring the structural information of biological tissue. Recently, many multi-contrast MRI super-resolution networks achieve good effects. However, most studies ignore the impact of the inappropriate foreground scale and patch size of multi-contrast MRI, which probably leads to inappropriate feature alignment. To tackle this problem, we propose the Flexible Alignment Super-Resolution Network (FASR-Net) for multi-contrast MRI Super-Resolution. The Flexible Alignment module of FASR-Net consists of two modules for feature alignment. 
 (1) The Single-Multi Pyramid Alignment(S-A) module solves the situation where low-resolution (LR) images and reference (Ref) images have different scales. (2) The Multi-Multi Pyramid Alignment(M-A) module solves the situation where LR and Ref images have the same scale. Besides, we propose the Cross-Hierarchical Progressive Fusion (CHPF) module aiming at fusing the features effectively, further improving the image quality. Compared with other state-of-the-art methods, FASR-net achieves the most competitive results on FastMRI and IXI datasets. Our code will be available at \href{https://github.com/yimingliu123/FASR-Net}{https://github.com/yimingliu123/FASR-Net}. 
\end{abstract}


\section{Introduction}
\label{sec:introduction}
Magnetic Resonance Imaging (MRI) is a non-invasive imaging modality that enables the observation of three-dimensional detailed anatomical images and plays a significant role in providing clear information about soft tissue structure. However, during acquiring magnetic resonance images, patients have to endure physical and psychological discomfort, including irritating noise and acute anxiety.  To make the patient feel cozier, technically, it will reduce the retention time that patients stay in the strong magnetic field at the expense of image quality. Super-Resolution reconstruction technology can improve the image quality without changing the hardware settings, which is extensively utilized as the post-processing tool to overcome the difficulty in obtaining high-resolution (HR) MRI scans~\cite{introduction1},\cite{introduction2}.

The research on image SR is divided into two categories: single-image Super-Resolution (SISR) and reference-based image Super-Resolution (Ref-SR). SISR~\cite{SAN},~\cite{SRCNN},~\cite{RCAN},~\cite{RDN},~\cite{ESRT},~\cite{kd1},~\cite{kd2} only adopts a single low-resolution (LR) image to recover HR images, which highly depends on the prior knowledge of training sets. Some transformer-based methods achieve fantastic results in SISR, such as~\cite{ESRT},~\cite{SWIN-IR}. Since the domain shift between training sets and test sets and the complementary details are inferred by training sets, as a result, SISR often produces blurry effects because the HR textural features cannot be effectively recovered in the reconstruction process. The texture information of medical images is crucial evidence for doctors' diagnoses. Therefore, SISR is not suitable for the medical image Super-Resolution.

\begin{figure}[htbp]
	\centering{\includegraphics[scale=0.25]{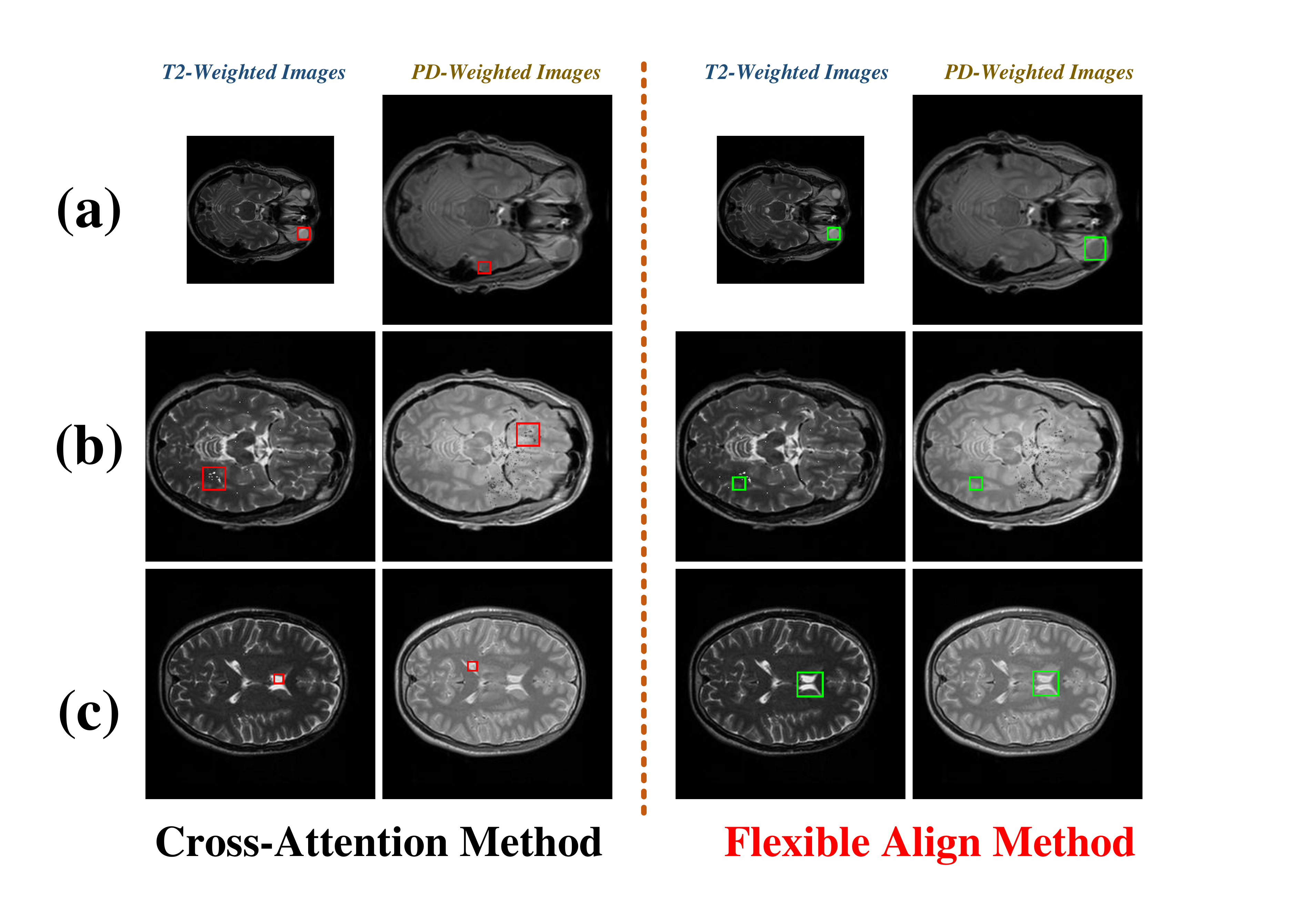}}
	\caption{\label{fig1}The comparison between Cross Attention method (CA) and our Flexible Align method (FA). FA contains two parts: S-A module and M-A module. Specifically, when the scale of LR and Ref images is different, S-A module will adjust the mismatch produced by CA, as shown in (a). M-A module solves the mismatch caused by the inappropriate patch size. If the patch size is too large, the similarity between patches will be dominated by background noise, which is illustrated in (b). When the patch size is too small, the insufficient semantics will lead to mismatch, which is denoted in (c). }
\end{figure}

Ref-SR~\cite{crossnet},~\cite{SSEN},~\cite{C2matching},~\cite{SRNTT},~\cite{TTSR},~\cite{MASA} adopts an additional high-resolution reference image as an auxiliary which transfers HR textures and details to the LR image in the process of SR. In clinical settings, MRI generates multi-contrast images for diagnosis together. Due to different settings, the appearances and functions of these images are widely divergent. However, these images can be used as complementary information for diagnosing the same anatomical structure. Generally, T1, T2, PD, and FS-PD weighted images are produced together in the acquisition of MRI. Clinically, PD-weighted images have shorter repetition and echo time than T2 weighted images~\cite{MINet}. Inspired by this, some Ref-SR based methods leverage the HR PD-weighted images to recover the HR T2 images from the LR T2 images.
For Ref-SR, some researches roughly complement LR image features with Ref image features through the plus or concatenation operation, where the improvement of the LR images quality is limited. Subsequently, a series of methods adopt deformable convolution. ~\cite{EDVR},~\cite{SRNTT}to fuse the Ref image features and LR image features. Existing state-of-the-art (SOTA) feature fusion methods subsequently unfold the image into patches and adopt Transformer-based cross-attention (CA) mechanism~\cite{SRNTT},~\cite{TTSR} to calculate the correlation between patches of LR and Ref images. These methods have verified that the feature alignment, which means matching valuable information of the patch from Ref images to the corresponding LR images, strongly impacts the reconstruction of HR images. 

Distinct from natural images, the color of MRI is sole and the object boundary is more ambiguous. Relevant experiments indicate that despite the existence of authentic high-frequency details in Ref images, the network cannot completely transform these details into HR images. We divide the MRI images into two parts: foreground and background. Specifically, the foreground contains some concerning tissue and texture which are important in SR. The background consists of some less important regions, such as the black region and the skeleton where the pixel information is nearly equal. Theoretically, the cross-attention (CA) methods only consider the search for the most relevant regions but ignore the variety of the scale of foreground. Through a large amount of experiments and observations, we find that the flexibility of patches has a significant effect on the feature alignment. Specifically, consider two cases:

1) When the LR and Ref image foregrounds are of different scales, the patch will contain different areas of the foreground. However, the foregrounds of the LR and Ref images have the same semantic information. The affinity of the semantic information will lead to the mismatch between LR and Ref image patches, as illustrated in \textcolor{cyan}{Fig. \ref{fig1}(a)}.  

2)  Assumed that the foreground scales of LR and Ref images are the same, the cross-attention (CA) method ignores the harmony between the patch size and the scale of the foreground. The fixed patch size barely adapts to the various scale of foregrounds. Therefore, the patch size is hard to fit the foreground scale. For example, if the size of the foreground is smaller than patch size, the patch will contain massive amounts of information, further interfering with the calculation of the correlation matrix, as illustrated in \textcolor{cyan}{Fig. \ref{fig1}(b)}. If the patch size is too small, different patches will mismatch due to the similarity of local features, as illustrated in \textcolor{cyan}{Fig. \ref{fig1}(c)}. 

In fact, scale diversity has been shown to be important in feature expression~\cite{pyramid} and image restoration~\cite{cai2022coarse},~\cite{liang2021high}. Small-scale features can provide more
complete semantic features, while large-scale feature maps can provide texture details. Based on this consideration, the core concept of the patch size can be illustrated as follows. (1) The patch should contain sufficient foreground information which contributes to the alignment. (2) In the meantime, disturbed background information is not expected too much in patch. To meet these demands, the receptive fields of patch should be adjustable. Therefore, we propose the Flexible Alignment (FA) module aiming at generating various patch size and receptive field to improve the precision of feature alignment.  Specifically, FA contains the Single-Multi Pyramid Alignment module (S-A) and the Multi-Multi Pyramid Alignment module (M-A) which respectively serves for the case I and case II. S-A leverages various receptive field to ensure the completeness of foreground information. M-A dynamically adjusts the patch size of LR and Ref images to escape from the influence of background. Additionally, we fuse the multi-scale features with the Cross-Hierarchical Progressive Fusion (CHPF) module, further improving the image quality. Furthermore, fourier loss function is introduced to optimize the model. Our contributions can be summarized as follows:

$\bullet$ We propose the FASR-Net to transform the textural information of high-resolution PD images into low-resolution T2 images and make the texture more realistic.

$\bullet$ Our model jointly combines the Multi-Multi Pyramid Alignment module (M-A) and the Single-Multi Pyramid Alignment module (S-A) to endow feature alignment with flexibility.   

$\bullet$ We introduce an effective feature fusion backbone Cross-Hierarchical Progressive Fusion (CHPF) which takes advantage of textural information and details of multi-scale features.

Our code will be available at \href{https://github.com/yimingliu123/FASR-Net}{FASR-Net}.

\section{Related Work}
\begin{figure*}[h]
	\centering{\includegraphics[scale=0.25]{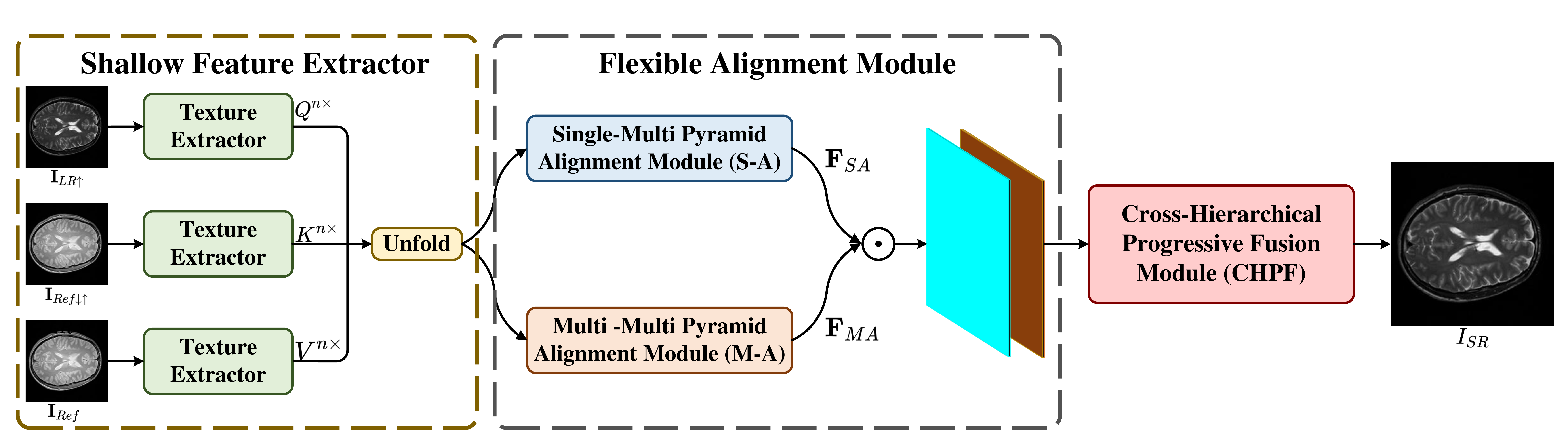}}
	\caption{\label{overview} The overview of RCFA-Net. RCFA-Net is composed of three parts: the shallow feature extractor, the Flexible Alignment module, and the Cross-Hierarchical Progressive Fusion module. $Q^{n\times}$, $K^{n\times}$, and $V^{n\times}$ are multi-scale shallow features extracted from $\mathbf{I}_{LR\uparrow}$, $\mathbf{I}_{Ref\downarrow\uparrow}$, and  $\mathbf{I}_{Ref}$. The S-A and M-A are two parts of the Flexible Alignment module introduced in Section III C. $\mathbf{F}_{SA}$ and $\mathbf{F}_{MA}$ are respectively aligned features produced by S-A and M-A module. After concatenating the features, the Cross-Hierarchical Progressive Fusion module is utilized to reconstruct the $\mathbf{I}_{SR}$.}
\end{figure*}

\subsection{Single Image Super-Resolution}
For the past few years, deep learning-based SISR methods have performed amazing performances. Some Coarse-to-Fine works~\cite{cai2022coarse},~\cite{liang2021high} have attractive results. Cai et al.~\cite{cai2022coarse} proposed a novel Transformer-based method, coarse-to-fine sparse Transformer (CST). Specifically, CST uses spectraaware screening mechanism (SASM) for coarse patch selecting. Then the selected patches are fed into spectra-aggregation hashing multi-head self-attention (SAH-MSA) for fine pixel clustering and self similarity capturing. Liang et al.~\cite{liang2021high} focus on speeding-up the high-resolution photorealistic I2IT tasks based on closed-form Laplacian pyramid decomposition and reconstruction. Except for Coarse-to-Fine method, Lu \emph{et al.} propose SRCNN~\cite{SRCNN}, which introduces deep convolutional neural networks in the field of image Super-Resolution. Thereafter, residual blocks~\cite{RDN} and attention mechanisms~\cite{chen2021cross},~\cite{RCAN},~\cite{SAN} are introduced to deepen the network. However, these approaches improve the image quality restrictedly since the difficulty in recovering high-frequency details. Christian \emph{et al.}~\cite{SRGAN} firstly adopt Generative Adversarial Network (GAN) in SR tasks. According to Christian, minimizing the mean squared loss (MSE) often lacks high-frequency details. Thus,~\cite{SRGAN} utilizes the perceptual loss which consists of content loss and adversarial loss. Wang \emph{et al.}~\cite{ESRGAN} further propose enhanced generator and discriminator obtaining more perceptually competitive results. Yan \emph{et al.}~\cite{yan2021fine} adopt FASRGAN to discriminate each pixel of real and fake images. Knowledge distillation framework is also introduced in SR tasks, such as~\cite{kd1},~\cite{kd2}. Recently, some Transformer-based networks have been applied to SISR tasks~\cite{ESRT},~\cite{SWIN-IR}. ESRT~\cite{ESRT} adopts a high-preserving block and lightweight transformer backbone, achieving satisfying results with low computational cost. SwinIR~\cite{SWIN-IR} adopts shifted-window-based self-attention mechanism in Swin Transformer~\cite{Swin}.

Although SISR approaches achieve marvelous results in the natural image domain, these methods are not suitable for medical images. The details of HR medical images, which are significant for diagnosis, are generated by networks but are not authentic. Therefore, the reasonable method for MRI Super-Resolution is complementing high-frequency information from additional HR images. Therefore, Ref-SR methods are prior to achieving believing methods for medical images. 	                 

\subsection{Reference-Based Image Super-Resolution}
Ref-SR adopts an additional high-resolution reference image to resolve low-resolution image. Compared with SISR, Ref-SR is more likely to harvest accurate textural information. One branch of Ref-SR methods is to align LR and Ref images. CrossNet~\cite{crossnet} adopts an end-to-end and fully-convolutional neural network with the optical flow estimator to align Ref and LR images. However, this approach depends on the alignment of Ref and LR images to a great extent. Additionally, the utilization of optical flow neglects the long-range dependencies. SSEN~\cite{SSEN} introduces a stack of deformable convolution~\cite{DC} layers, enlarging the receptive field of Ref images. $C^2$-matching~\cite{C2matching} introduces the contrastive correspondence network and teacher-student correlation distillation to align images on pixel level. However, because the restoration module of $C^2$-matching contains only simple residual blocks, the misalignment between images will drastically destroy the performance of this method. 

Another mainstream of Ref-SR approaches is based on patch matching~\cite{patchmatch}. SRNTT~\cite{SRNTT} adopts cross-attention mechanism to achieve patch matching, which endows LR images with HR details by transferring textural information from Ref images according to the correlation. Further, TTSR~\cite{TTSR} retains the idea of cross-attention and introduces  soft-attention module which subsequently computes the relevance between original and swapped features and feeds all swapped features with different weights into the main network. MASA~\cite{MASA} takes the potential enormous difference, such as color and luminance distribution, into consideration and reduces the computational cost. Cao \emph{et al.}~\cite{cao2022reference} combine deformable attention with cross-attention mechanism, further improving the performance of Res-SR in exchange for the sacrifice of the computational cost.

\section{Method}
\subsection{Overview}
The Ref-SR methods aiming at transforming details of high-resolution reference (Ref) images into low-resolution (LR) images have achieved fantastic results recently. The details recovered by the Ref-SR methods are more reliable than the SISR methods.
Clinically, to observe a tissue extensively, a series of multi-contrast images will be produced together during the acquisition of MRI. Therefore, an intuitive thought is that the low-cost images (PD) can be used as references to offer helpful detail information for high-cost images (T2).  Based on this consideration, we propose a novel Multi-Contrast Flexible Alignment Super-Resolution Network (FASR-Net) for MRI. The architecture of FASR-Net is shown in \textcolor{cyan}{Fig.~\ref{overview}}.  $\mathbf{I}_{LR\uparrow}$ and  $\mathbf{I}_{Ref}$  represent upsampled T2 images and PD images respectively. We sequentially apply downsampling and upsampling with the same factor $4\times$ on PD images to obtain $\mathbf{I}_{Ref\downarrow\uparrow}$.

Functionally, the FASR-Net can be roughly divided into three parts: the shallow feature extractor, the Flexible Alignment (FA) module and the Cross-Hierarchical Progressive Fusion (CHPF) module. Specifically, the shallow feature extractor aims at obtaining robust semantic features. FA module which is composed of M-A and S-A serves for feature alignment. After the feature alignment, we leverage CHPF to fuse the aligned features.

\subsection{Texture Extractor}
The FASR-Net doesn't adopt transformer-based methods directly~\cite{VIT1},\cite{VIT2}, \cite{VIT3},\cite{Swin},\cite{boat},~\cite{pyramid},~\cite{ESRGAN},~\cite{SRGAN} which unfold the raw image to patches and feed into patch-embedding .
Instead, the raw image patches are first sent into a CNN-based backbone for shallow feature extraction. There are two main reasons: (1) In the following alignment process, not each pixel of the Ref image can exactly correspond to the local part of the LR image, so the use of shallow features can compensate for this pixel-by-pixel accuracy and make them semantically align with each other.
(2) Transformer focuses too much on feature interaction on a global level, but ignores the importance of local neighborhoods. Therefore, CNN based network~\cite{VGG19} is selected as the texture extractor to make up for the feature translation invariance that transformer block in FA module. In addition, the texture extraction part is relatively independent, which means a structure with a high degree of freedom. We can replace the backbone as needed. Several well-known backbones were analyzed in ablation experiments.

\begin{figure}[]
	\centering{\includegraphics[scale=0.16]{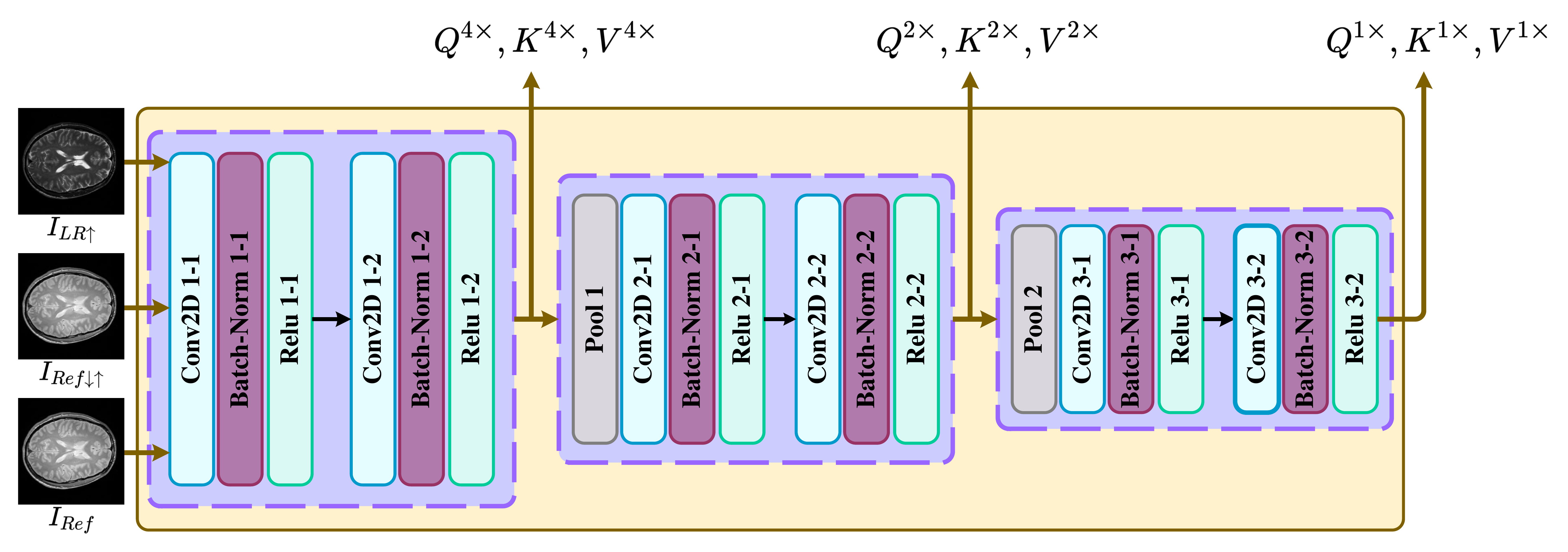}}
	\caption{\label{FE} We adopt a pretrained VGG-19 network as the texture extractor. $4\times$ feature, $2\times$ feature and $1\times$ feature are respectively extracted from Relu 1-2, Relu 2-2 and Relu 3-2. Note that the parameters of texture extractor will be finetuned in the process of training.}
\end{figure}

\subsection{Flexible Alignment Module}
Feature alignment is to find the most corresponding patches in Ref and LR images. However, the selection of patch size greatly affects the results of feature alignment. It is mainly reflected in two aspects: (1) If the sizes of the foreground are not the same in LR and Ref images, it will lead to mismatch due to the fixed size of patch. (2) Assuming that the sizes of the foreground are the same in both LR and Ref images, the inappropriate patch size will lead to interference caused by unnecessary background or inaccurate alignment due to insufficient foreground information. The FA module can be divided into two sub-modules, S-A solves the problems existing in aspect (1) by using multi-scale features as inputs. M-A leverages dynamical patch size to solve the problems existing in aspect (2).

\subsubsection{The Single-Multi Pyramid Alignment module (S-A) for different image scale}

\begin{figure}[ht]
	\centering{\includegraphics[scale=0.20]{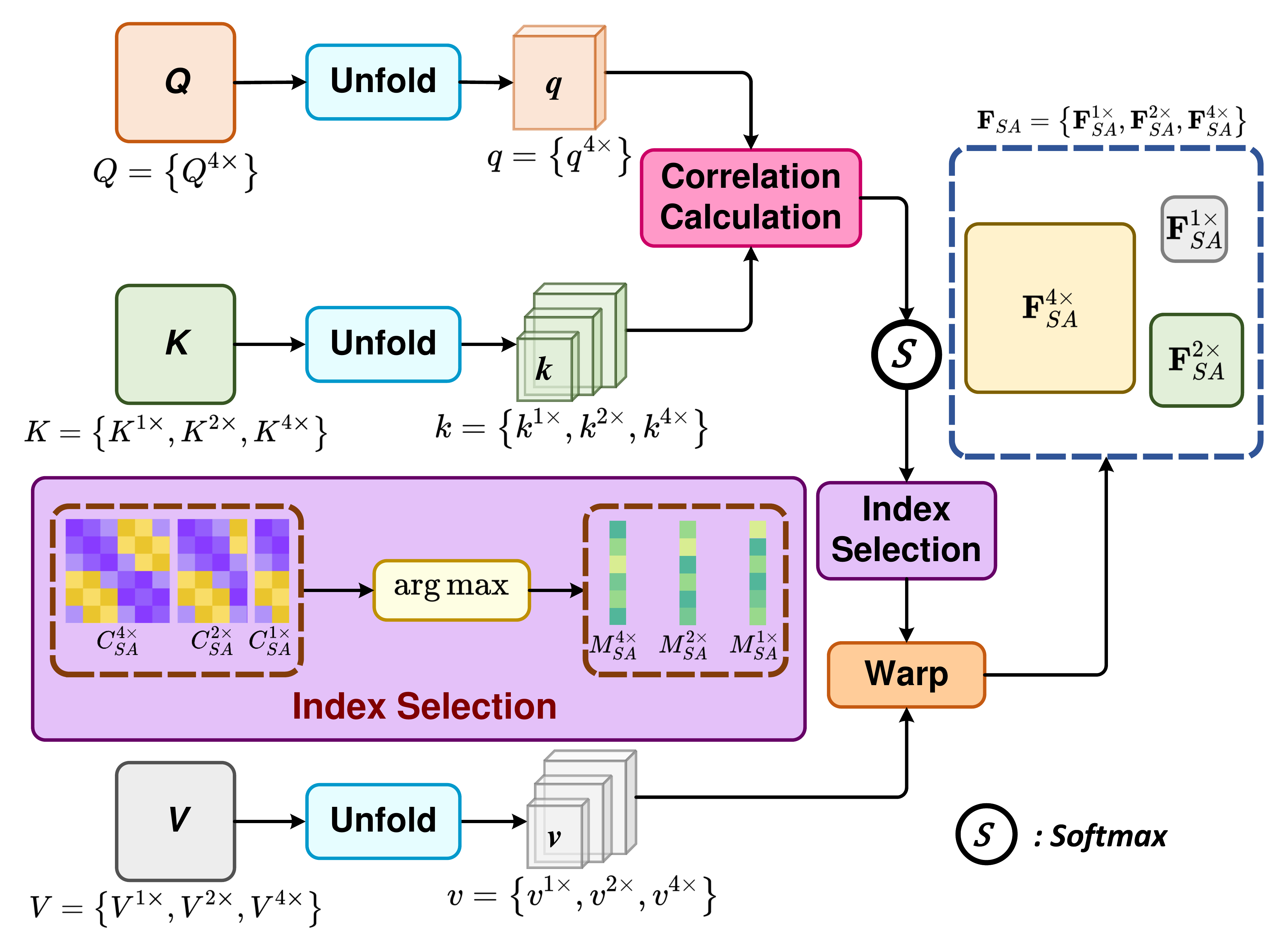}}
	\caption{\label{SM} The architecture of S-A in detail. The process of index selection is shown in the purple rectangle. The input $Q^{4\times}$, $K^{n\times}$, and $V^{n\times}$ are fed into word embedding to match the shape. Note that word embedding contains both unfold and linear layer. Through the correlation calculation between $q^{4\times}$ and $k^{n\times}$, we can obtain $C_{SA}^{n\times}$. Then $k^{n\times}$ will warp with the indication of corresponding $M_{SA}^{n\times}$, which is implemented by the index selection of $C_{SA}^{n\times}$.} 
\end{figure}

 The process of patch-matching depends on the similarity scores. However, some mismatched patches of foreground will gain high similarity scores since the semantical similarity, as seen in \textcolor{cyan}{Fig.~\ref{fig1}(a)}. To solve this problem, we propose the Single-Multi Pyramid Alignment module to endow $\mathbf{I}_{Ref\downarrow\uparrow}$ with flexible receptive fields. In S-A module, the $\mathbf{I}_{LR\uparrow}$ and $\mathbf{I}_{Ref\downarrow\uparrow}$ are of different image scales. The operation takes full advantage of the patches with different scales, as illustrated in \textcolor{cyan}{Fig.~\ref{SM}}.

The S-A module generates $q^{4\times}$  $k^{n\times}$, and $v^{n\times}$ during word embedding, the process can be expressed as:
\begin{equation}
	q^{4\times} = \mathcal{F}_{WE-q}(Q^{4\times})
	\label{eq:7}
\end{equation}
\begin{equation}
	k^{n\times} = \mathcal{F}_{WE-k}(K^{n\times})	
	\label{eq:8}
\end{equation}
\begin{equation}
	v^{n\times} = \mathcal{F}_{WE-v}(V^{n\times})	
	\label{eq:9}
\end{equation}
where $n=\left\{1,2,4\right\}$, which represents the factor $1\times$, $2\times$ and $4\times$ of texture feature extracted from different depth of the backbone, as can be seen in \textcolor{cyan}{Fig.~\ref{FE}}.

The alignment process is in the purple rectangle of \textcolor{cyan}{Fig.~\ref{SM}}. Inspired by the cross-attention mechanism~\cite{cross}, we generate $C_{SA}^{n\times}$ matrix which represents the correlation between different embedded patches from  $q^{4\times}$ and $k^{n\times}$, as illustrated in Eq.~\eqref{eq:10}.
\begin{equation}
	C_{SA}^{n\times} = \pmb {Softmax}( \langle q^{4\times},k^{n\times} \rangle,  dim=1)
	\label{eq:10}
\end{equation}

After calculating the similar scores $C_{SA}^{n\times}(i)$, we create $M_{SA}^{n\times}(i)$ through $C_{SA}^{n\times}$ matrix, where we only save the index of max weight position in each row, as illustrated in Eq.~\eqref{eq11}. Then we use $M_{SA}^{n\times}(i)$ matrix to warp the $V^{n\times}$, as seen in Eq.~\eqref{eq12}. The warp operation means the embedding patches with important detail information on the Ref image are preserved and those with few reference value are obliterated. The output of S-A module $\mathbf{F}_{SA}^{n\times}$ is summarized in Eq.~\eqref{eq13}.
\begin{equation}
	M_{SA}^{n\times}(i)=\arg\max\limits_{j}{{C}_{SA}^{n\times}(i,j)}
	\label{eq11}
\end{equation}
\begin{equation}
	\mathbf{F}_{SA}^{n\times}=v^{n\times}(M_{SA}^{n\times}(i))
	\label{eq12}
\end{equation}
\begin{equation}
	\mathbf{F}_{SA}=\left\{\mathbf{F}_{SA}^{1\times},\mathbf{F}_{SA}^{2\times}, \mathbf{F}_{SA}^{4\times}\right\}
	\label{eq13}
\end{equation}
where $(i,j)$ denotes the index of 2D tensors (${C}_{SA}^{n\times}(i,j)$), $(i)$ represents the row index of 2D tensors ($v^{n\times}$) and the index of 1D tensors ($M_{SA}^{n\times}$).

\subsubsection{The Multi-Multi Pyramid Alignment (M-A) module for the same image scale}

\begin{figure}[ht]
	\centering{\includegraphics[scale=0.20]{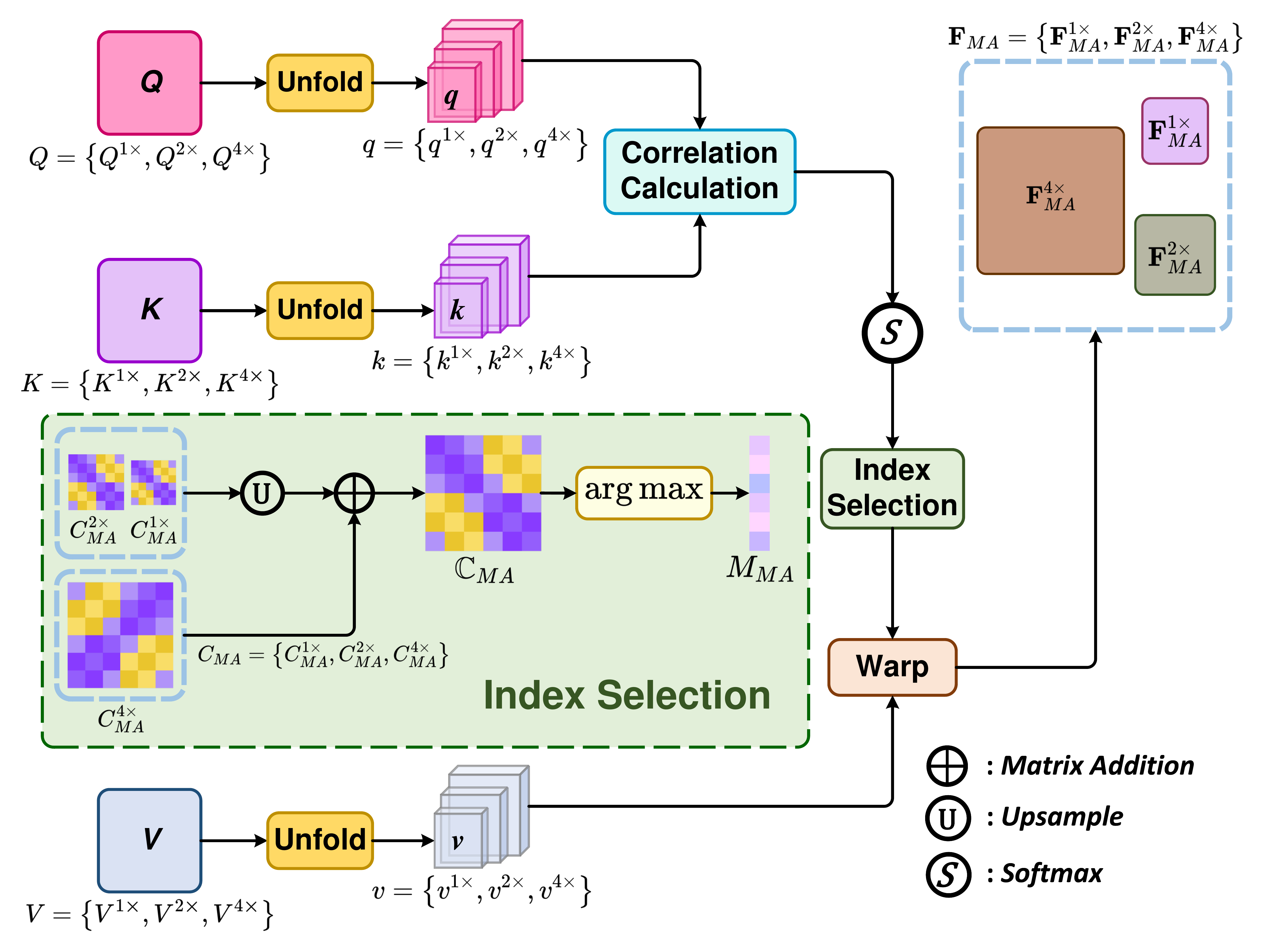}}
	\caption{\label{MM} The architecture of M-A in detail. The process of index selection is shown in the green rectangle. The inputs $Q^{n\times}$, $K^{n\times}$ and $V^{n\times}$ are unfolded to multi-scale two-dimensional tensors (Note that the batch dimension is discarded from notations for simplicity). Because the shape of matrix ($C_{MA}^{n\times}$) row is different, we upsample the matrixes to match the shape. Subsequently, all muti-scale $v^{n\times}$ warp with the order of $M_{MA}$.}
\end{figure}

The invariance of the patch size limits the performance of feature alignment. When the proportion of foreground in the patch is small, even if $\mathbf{I}_{LR\uparrow}$ and $\mathbf{I}_{Ref\downarrow\uparrow}$ have high similarity in the foreground, it will still fail to match due to the interference of different backgrounds or insufficient foreground information, as seen in \textcolor{cyan}{Fig.~\ref{fig1}(b)} and \textcolor{cyan}{Fig.~\ref{fig1}(c)}. To ensure the appropriate proportion of foreground in the patch, we introduce M-A module, where the texture features of $\mathbf{I}_{LR\uparrow}$ and $\mathbf{I}_{Ref\downarrow\uparrow}$ have the same image scale, as shown in \textcolor{cyan}{Fig.~\ref{MM}}.

$M-A$ receives query $Q^{n\times}\in\mathbb{C}^{f_{n\times}\times w \times h}$, key $K^{n\times}\in\mathbb{C}^{f_{n\times}\times w \times h}$ and
$V^{n\times}\in\mathbb{C}^{f_{n\times}\times w \times h}$ as the inputs. Subsequently, we unfold the inputs into two-dimensional tensors $q^{n\times}\in\mathbb{C}^{f_{n\times}\times wh}$, $k^{n\times}\in\mathbb{C}^{f_{n\times}\times wh}$ and $v^{n\times}\in\mathbb{C}^{f_{n\times}\times wh}$. 

Eq.~\eqref{eq:17} performs the similarity response in the M-A module. The response $C_{MA}^{n\times}$ matrixes can form a new pyramid structure. To match the different shapes of $C_{MA}^{n\times}$, we apply upsampling to $C_{MA}^{n\times}$, as illustrated in Eq.~\eqref{eq:18} where $\mathcal{U}$ represents upsampling.
\begin{equation}
	C_{MA}^{n\times} = \pmb {Softmax}( \langle q^{n\times},k^{n\times} \rangle,  dim=1)
	\label{eq:17}
\end{equation}
\begin{equation}
	\mathbb{C}_{MA} = \mathcal{U}(C^{1\times}_{MA}) + \mathcal{U}(C^{2\times}_{MA}) + C^{4\times}_{MA}
	\label{eq:18}
\end{equation}

Similar to S-A, we sort the rows of $\mathbb{C}_{MA}$ matrix and generate $M_{MA}(i)$ by preserving the position with max weights. Thus, we acquire ${F}_{MA}$.

\subsection{The Cross-Hierarchical Progressive Fusion module (CHPF)}
\begin{figure*}[ht]
	\centering{\includegraphics[scale=0.2]{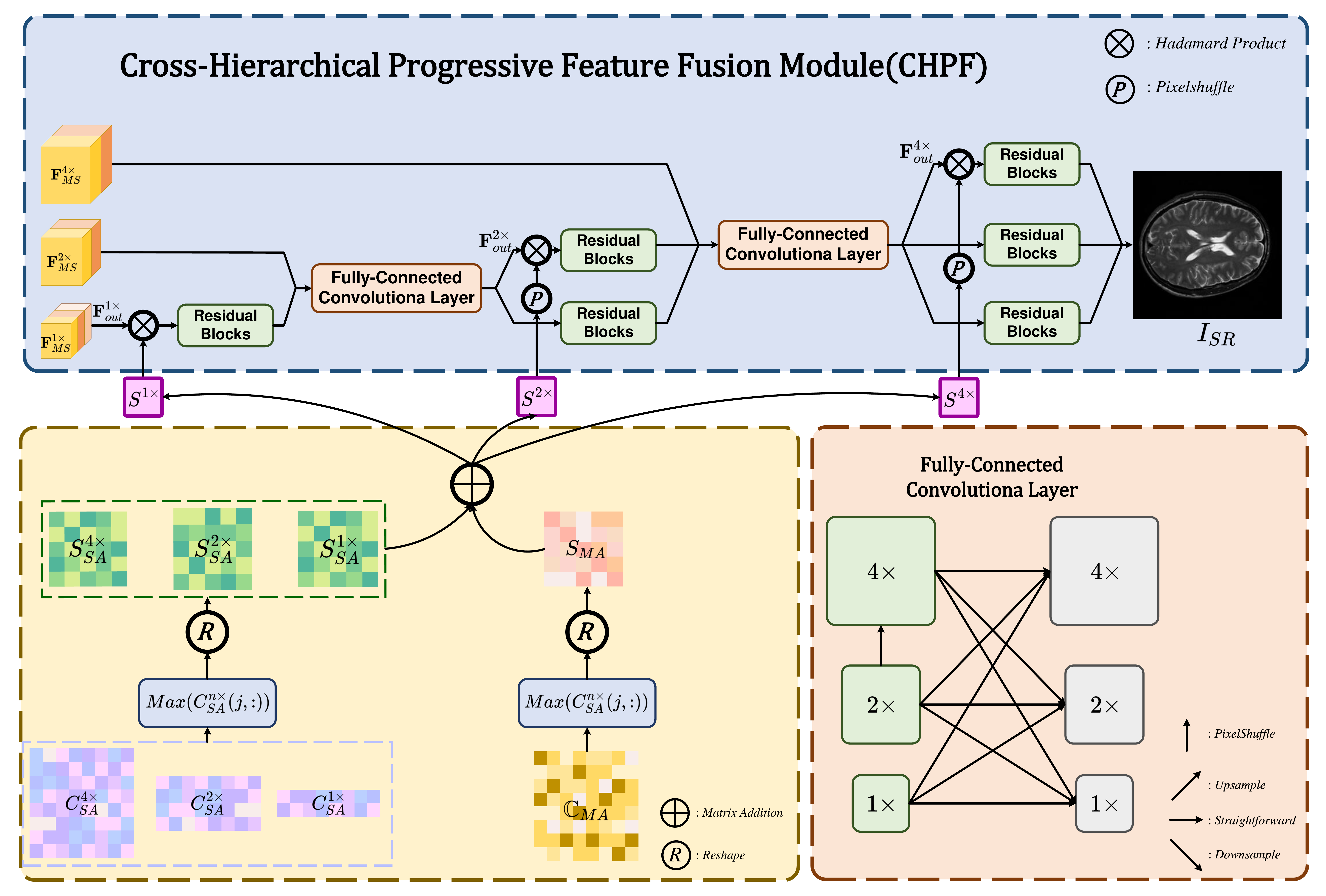}}
	\caption{\label{CHPF} The backbone of CHPF is shown in the top rectangle. To take full advantage of multi-scale features, $1\times$, $2\times$ and $4\times$ features will fuse in CHPF. To match the shapes, we adopt the fully-connected convolutional layer to merge multi-scale features, as illustrated at the bottom-right corner. Note that we also introduce soft-attention mechanism, which means $S^{n\times}$ indicates the level of transferred texture information. The calculation of $S^{n\times}$ is shown in the yellow rectangle.}
\end{figure*}

After the Flexible Alignment (FA) module, the texture feature has been completely aligned in various scales thanks to the pyramid model structure. In the final feature fusion stage, we propose a Cross-Hierarchical Progressive Fusion module (CHPF) to fuse the features of different scales and restore $\mathbf{I}_{SR}$. As shown in \textcolor{cyan}{Fig.~\ref{CHPF}}, the input of the CHPF module is $\mathbf{F}_{MS}^{n\times}$ generated by  Eq. \eqref{eq:23}. $\mathbf{F}_{MA}^{n\times}$ and $\mathbf{F}_{SA}^{n\times}$ represent the outputs of FA module. In order to leverage muti-scale features (include $1\times$, $2\times$ and $4\times$ features), $\mathbf{F}_{MS}^{n\times}$ of various scale will map to the same size by up/down sampling in fully-connection convolutional layer. In addition, the fully-connection convolutional layer will also fuse the $\mathbf{F}_{MS}^{n\times}$ by $1\times1$ convolution, denoted by straightforward.
\begin{equation}
	\mathbf{F}_{in}^{n\times}=Concat(\mathbf{F}_{MA}^{n\times},\mathbf{F}_{SA}^{n\times})=\left\{\mathbf{F}_{in}^{1\times},\mathbf{F}_{in}^{2\times}, \mathbf{F}_{in}^{4\times}\right\}
	\label{eq:23}
\end{equation}

In fact, each scale of $\mathbf{F}_{out}^{n\times}$ has transferred texture information with different levels. During feature fusion, we should treat them discriminatively. Therefore, each scale of $\mathbf{F}^{n\times}$ will be modulated by soft-attention mechanism, as can be seen in Eq. \eqref{eq:27}. Soft-weight matrixes ($S^{n\times}$) are generated from both the S-A and the M-A modules, as shown in Eq. \eqref{eq:25}-\eqref{eq:28}.

\begin{equation}
	S_{SA}^{n\times}(i)=\max\limits_{j}{{C}_{SA}(i,j)}	
	\label{eq:25}
\end{equation}
\begin{equation}
	S_{MA}(i)=\max\limits_{j}{\mathbb{C}_{MA}(i,j)}		
	\label{eq:26}
\end{equation}
\begin{equation}		
	S^{n\times}=S_{SA}^{n\times}+S_{MA}
	\label{eq:28}
\end{equation}
\begin{equation}		
	\mathbf{F}^{n\times}=\mathbf{F}_{out}^{n\times} \otimes S^{n\times}
	\label{eq:27}
\end{equation}
where $\otimes$ represents matrix multiplication.

\subsection{Loss}

\subsubsection{L1 Loss}
To distinguish the differences between the high-resolution reconstructed image ($\mathbf{I}_{SR}$) and ground truth ($\mathbf{I}_{HR}$), a practical method is calculating the pixel-level difference between the two images. This method can guide FASR-Net to pay more attention on the details of the image, such as texture, color, and blurred information. Many previous experiments have proved its effectiveness~\cite{l11},~\cite{l12}. We choose L1 loss to act as a pixel-level loss in FASR-Net. It acts as an essential part of the total loss. For details of the L1 loss, see Eq.~\eqref{eq:29}. The total loss can be seen in Eq.~\eqref{eq:32}. Where $\mathbf{I}_{HR}$ represents ground truth, and $\mathbf{I}_{SR}$ represents the image that was repaired.
\begin{equation}
	{\mathcal{L}_1} = |\mathbf{I}_{SR} - \mathbf{I}_{HR}|
	\label{eq:29}
\end{equation}

\subsubsection{Ssim Loss}
However, using the L1 loss alone is limited. It will focus only on local detail but ignore the global information. Besides that, it can also lead to artifacts. For instance, considering two identical images offset by one pixel from each other, although they are similar in perception, the results of L1 loss will be very different. Based on this consideration, the FASR-Net will calculate the differences in terms of the total structural similarity. So we incorporate the ssim loss into the total-loss function to loosen the restrictions of using L1 loss alone. This loss calculates the structure similarity and other major sensory indicators such as brightness and contrast between ground truth and the repaired image, which is proved to be beneficial to human subjective perception. The process is shown in Eq.~\eqref{eq:30} and Eq.~\eqref{eq:31}.
\begin{equation}
	{\mathcal{L}_{ssim}} = {1-ssim(\mathbf{I}_{SR}, \mathbf{I}_{HR})}
	\label{eq:30}
\end{equation}
\begin{equation}
	ssim(x,y) = \frac{(2\mu_{x}\mu_{y}+c_{1})(2\sigma_{xy}+c_{2})}{(\mu_{x}^{2}+\mu_{y}^{2}+c_{1})(\sigma_{x}^{2}+\sigma_{y}^{2}+c_{2})}
	\label{eq:31}
\end{equation}
where, $\mu_{x}$ and $\mu_{y}$ are the means of $x$ and $y$, respectively. $\sigma_{x}^2$ and $\sigma_{y}^2$ are the variances of $x$ and $y$. $\sigma_{xy}$ is the covariance. Additionally, $c_{1}$ and $c_{2}$ are the constants used to maintain stability.

\subsubsection{Fourier Transform Loss}
Since the purpose of image Super-Resolution is more critical to the reconstruction of the lost high-frequency components, it is crucial to reduce the variance in the frequency domain. Consequently, we apply a Frequency Reconstruction (FR) loss, as can be seen in Eq. ~\eqref{eq:32}.
\begin{equation}
	{\mathcal{L}_{FR}} =  {\left\| {F\left( \mathbf{I}_{SR} \right) - F\left( \mathbf{I}_{HR} \right)} \right\|_1}
	\label{eq:32}
\end{equation}
$F$ denotes the fast Fourier transform (FFT) that transfers the image signal to the frequency domain. The final loss function for training our network is determined as follows:

\begin{equation}
	{\mathcal{L}_{total}} = {\mathcal{L}_1} + {\lambda_{1}\mathcal{L}_{ssim}} + {\lambda_{2}\mathcal{L}_{FR}}
	\label{eq:33}
\end{equation}
where,$\lambda_{1}$ and $\lambda_{2}$ are the hyper-parameters used to balance the L1, ssim and FR loss.

\section{Experiments}

\subsection{Datasets}
\subsubsection{IXI}
The IXI dataset~\cite{IXI} is a collection of MR images from $578$ patients, including T1, T2, and PD-weighted images and others. The T1, T2, and PD images obtained from the same region each contain special textural details. If every single image can be utilized properly, a higher resolution image can be generated. 

Our experiment uses downsampled T2 images as inputs, original T2 images as ground truth (GT), and corresponding PD images as reference. Specifically, the original size of both T2 and PD images in the IXI dataset are $256\times256\times3$. The LR images are created by downsampling. For example, to achieve the $4\times$ scale of SR results, its size should be $64\times64\times3$. Before training, all images are normalized over the range of $[-1, 1]$. We filter out $14000$ and $1000$ pairs of T2 and PD-weighted images for training and validation, respectively.

\subsubsection{FastMRI}
The FastMRI dataset~\cite{fastmri} contains four types of data from MRI acquisitions of knees and brains, aiming to connect the data science and the MRI research communities. The Knee MRI subset embodies raw data from more than $1,500$ fully sampled knee MRIs obtained on $3$ and $1.5$ Tesla magnets, and DICOM images from 10,000 clinical knees MRIs also obtained at $3$ or $1.5$ Tesla. The Brain MRI collects  $6,970$ fully sampled brain MRIs obtained on $3$ and $1.5$ Tesla magnets. It must be mentioned that the shape of original images varies for different patients.
We filter out $600$ and $58$ pairs of PD and FS-PDWI (Fat Suppression Proton Density-Weighted images which have been aligned to 256x256x3 for training and validation. The other configuration is in accord with the IXI training protocol.

\begin{figure*}[htbp] 
	\centering{
		\includegraphics[scale=0.22]{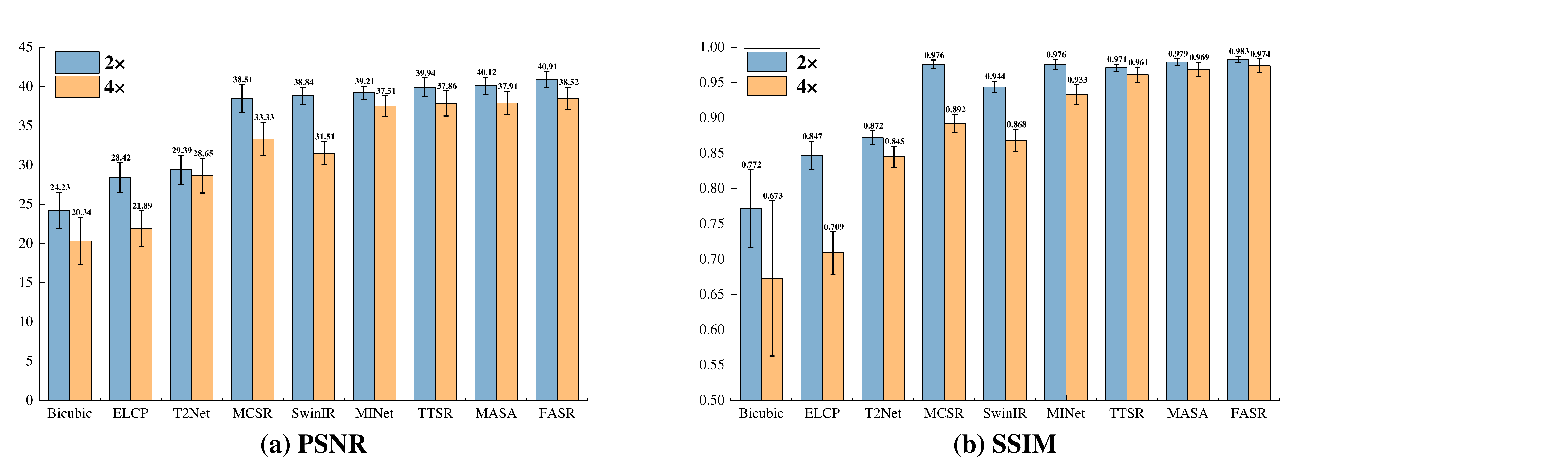}
	}
	\caption{ \label{fig:ixi} The quantitative metrics of IXI dataset under $2\times$ and $4\times$ enlargement.}
\end{figure*}

\subsection{Comparion with the SOTA}
To demonstrate that our proposed model can achieve better SR results, we compare the experimental results on IXI and FastMRI, respectively. We employ PSNR and SSIM as criteria to evaluate the effectiveness of our method. 

\subsubsection{IXI Dataset}
We compare our results with several previous works on the IXI dataset, including ELCP\cite{ELCP}, T2Net\cite{T2Net}, MCSR\cite{MCSR}, SwinIR\cite{SWIN-IR}, MINet\cite{MINet}, TTSR\cite{TTSR} and MASA\cite{MASA}. \textcolor{cyan}{Fig.~\ref{fig:ixi}} reports the metrics scores with IXI dataset under $2\times$ and $4\times$ enlargement. As can be seen, our approach achieves the best performance compared with the existing method. ELCP~\cite{ELCP} uses a convolutional neural network to integrate diverse SR results from individual models, combining multiple GAN models trained on the different image prior datasets. However, it can be found that the effect is not ideal in extensive multiple Super-Resolution repairs. T2Net~\cite{T2Net} combines reconstruction and Super-Resolution and encourages joint feature learning between the two tasks. However, no reference information was used compared  with us, so the repair effect was not optimistic. SwinIR~\cite{SWIN-IR} adopts swin-transformer as feature extractor and utilize a common decoder to obtain SR results. Due to the robust feature extracted by swin-transformer, this method achieves best results among the SISR methods. Notably, the SISR methods are far less effective than the Ref-SR models. MCSR~\cite{MCSR} exploits a contrasting HR image of a different modality as a reference. A CNN-based MCSR-processing step is proposed to  improve the quality of the reconstructed further. However, there is still some gap compared with our model due to the lack of accurate feature fusion for reference details. MINet~\cite{MINet} learns a hierarchical feature representation from multiple convolutional stages for each different-contrast image, and good results are obtained with the help of reference images. TTSR~\cite{TTSR} and MASA~\cite{MASA} introduces align module. Therefore, these methods obtain more effective results.

\begin{figure*}[] 
	\centering{
		\includegraphics[scale=0.28]{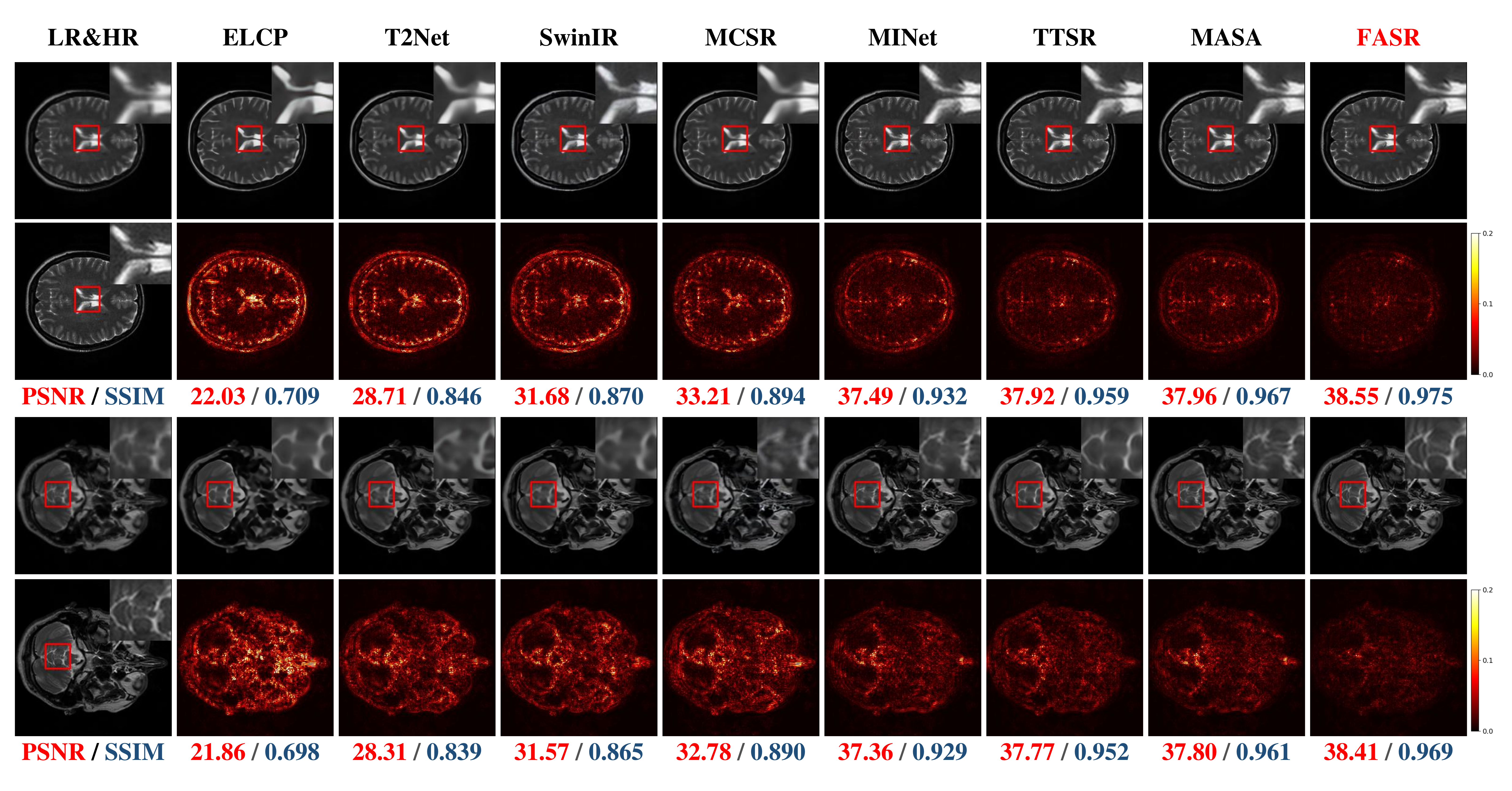}
	}
	\caption{ \label{fig:IXI_v} Visual comparison of the output and residual maps from different model on the IXI dataset under $4\times$ enlargement. The residual map measures the absolute value of pixel-value disparities between the output of the model and corresponding ground truth. LR represents low-resolution images which generated by downsampling. HR represents the ground truth.}
\end{figure*}

In \textcolor{cyan}{Fig.~\ref{fig:IXI_v}}, we use the $4\times$ scale output $\mathbf{I}_{SR}$ and the corresponding residual maps to show our model's best visual effect. The residual maps, also named error maps, represent the absolute pixel-value differences between the output results from each method and the ground truth of the T2-weighted HR image. The smaller the difference, the closer the two images are, which means the better reconstruction performance the model achieves. 

\subsubsection{FastMRI Dataset}
We compare our results with several previous works on the FastMRI dataset, including  EDSR\cite{EDSR}, MCSR\cite{MCSR}, MINet\cite{MINet}, TTSR\cite{TTSR}, MASA\cite{MASA} and SwinIR\cite{SWIN-IR}.

\begingroup
\setlength{\tabcolsep}{10pt} 
\renewcommand{\arraystretch}{1.5} 
\begin{table}[htbp]	
	\centering
	\caption{\label{tab:fastmri} Quantitive Comparision on the FastMRI dataset}
	\scalebox{0.9}{
		\begin{tabu}{ccc|cccc|cccc}
			\tabucline[1.5pt]{-}
			\multicolumn{3}{l|}{Scale}                & \multicolumn{4}{c|}{2$\times$}                               & \multicolumn{4}{c}{4$\times$}                                  \\ \hline
			\multicolumn{3}{l|}{Metrics}              & \multicolumn{2}{c}{PSNR} & \multicolumn{2}{c|}{SSIM} & \multicolumn{2}{c}{PSNR}   & \multicolumn{2}{c}{SSIM}  \\ \hline \hline
			\multicolumn{3}{l|}{Bicubic}              & \multicolumn{2}{l}{23.141}     & \multicolumn{2}{l|}{0.714}     & \multicolumn{2}{l}{19.202} & \multicolumn{2}{l}{0.645} \\
			\multicolumn{3}{l|}{EDSR (2017,CVPR)~\cite{EDSR}}     & \multicolumn{2}{l}{30.124}     & \multicolumn{2}{l|}{0.851}     & \multicolumn{2}{l}{26.719} & \multicolumn{2}{l}{0.725} \\
			\multicolumn{3}{l|}{MCSR (2020, TMI)~\cite{MCSR}}     & \multicolumn{2}{l}{35.021}     & \multicolumn{2}{l|}{0.897}     & \multicolumn{2}{l}{29.542} & \multicolumn{2}{l}{0.816} \\
			\multicolumn{3}{l|}{TTSR (2020, CVPR)~\cite{TTSR}}  & \multicolumn{2}{l}{35.712}     & \multicolumn{2}{l|}{0.909}     & \multicolumn{2}{l}{32.018}       & \multicolumn{2}{l}{\textbf{0.880}}      \\
			\multicolumn{3}{l|}{MINet (2021, MICCAI)~\cite{MINet}} & \multicolumn{2}{l}{35.917}     & \multicolumn{2}{l|}{0.926}     & \multicolumn{2}{l}{31.974} & \multicolumn{2}{l}{0.876} \\
			\multicolumn{3}{l|}{MASA (2021, CVPR)~\cite{MASA}}    & \multicolumn{2}{l}{36.817}     & \multicolumn{2}{l|}{0.934}     & \multicolumn{2}{l}{32.965} & \multicolumn{2}{l}{0.842} \\
			\multicolumn{3}{l|}{SwinIR (2021, ICCV)~\cite{SWIN-IR}}  & \multicolumn{2}{l}{37.729}     & \multicolumn{2}{l|}{0.957}     & \multicolumn{2}{l}{32.971} & \multicolumn{2}{l}{0.849} \\
			\multicolumn{3}{l|}{FASR\textbf{(Ours)}}          & \multicolumn{2}{l}{\textbf{{38.231}}}     & \multicolumn{2}{l|}{\textbf{{0.964}}}     & \multicolumn{2}{l}{\textbf{{33.403}}}       & \multicolumn{2}{l}{{0.861}}     \\  \tabucline[1.5pt]{-}
	\end{tabu}}
	\label{tab:addlabel}%
\end{table}

The corresponding visual effects are displayed in \textcolor{cyan}{Fig. \ref{fig:fastmri}}, the differential effect between our result and ground truth can be easily found in the residual map. The lighter the color is in the residual map, the better the effect is.

\begin{figure*}[ht]
	\centering{
		\includegraphics[scale=0.3]{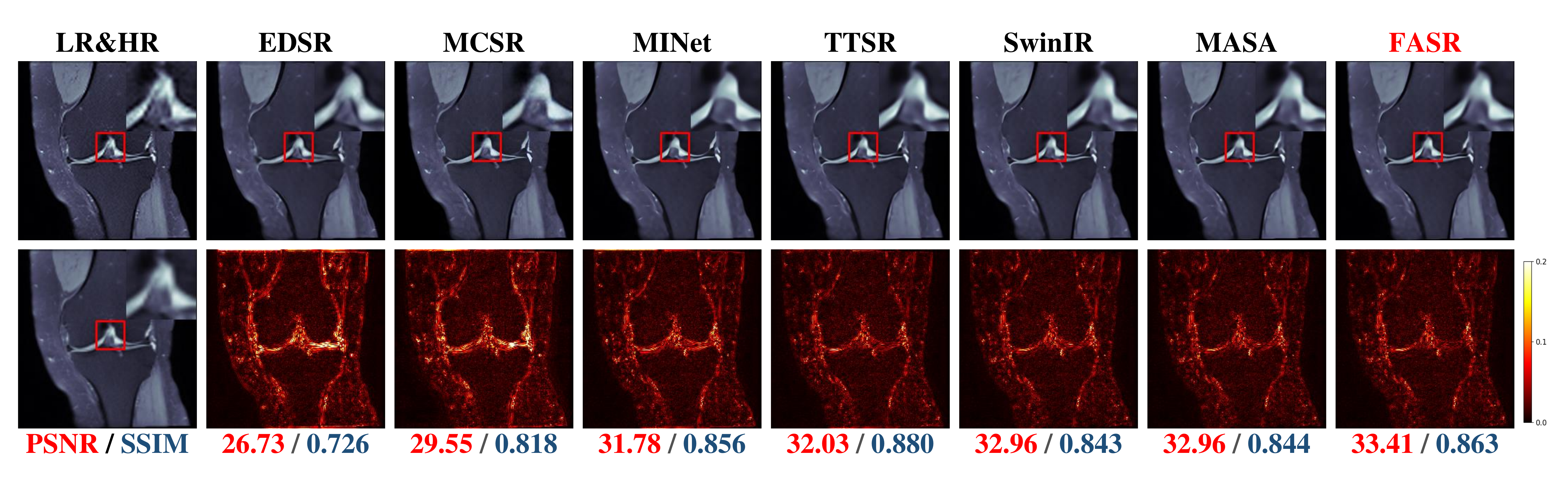}
	}
	\caption{\label{fig:fastmri}  Visual comparison of the output and residual maps from different models on the FastMRI dataset under 4$\times$ enlargement.}
\end{figure*}

\begin{figure*}[ht]
	\centering{\includegraphics[scale=0.45]{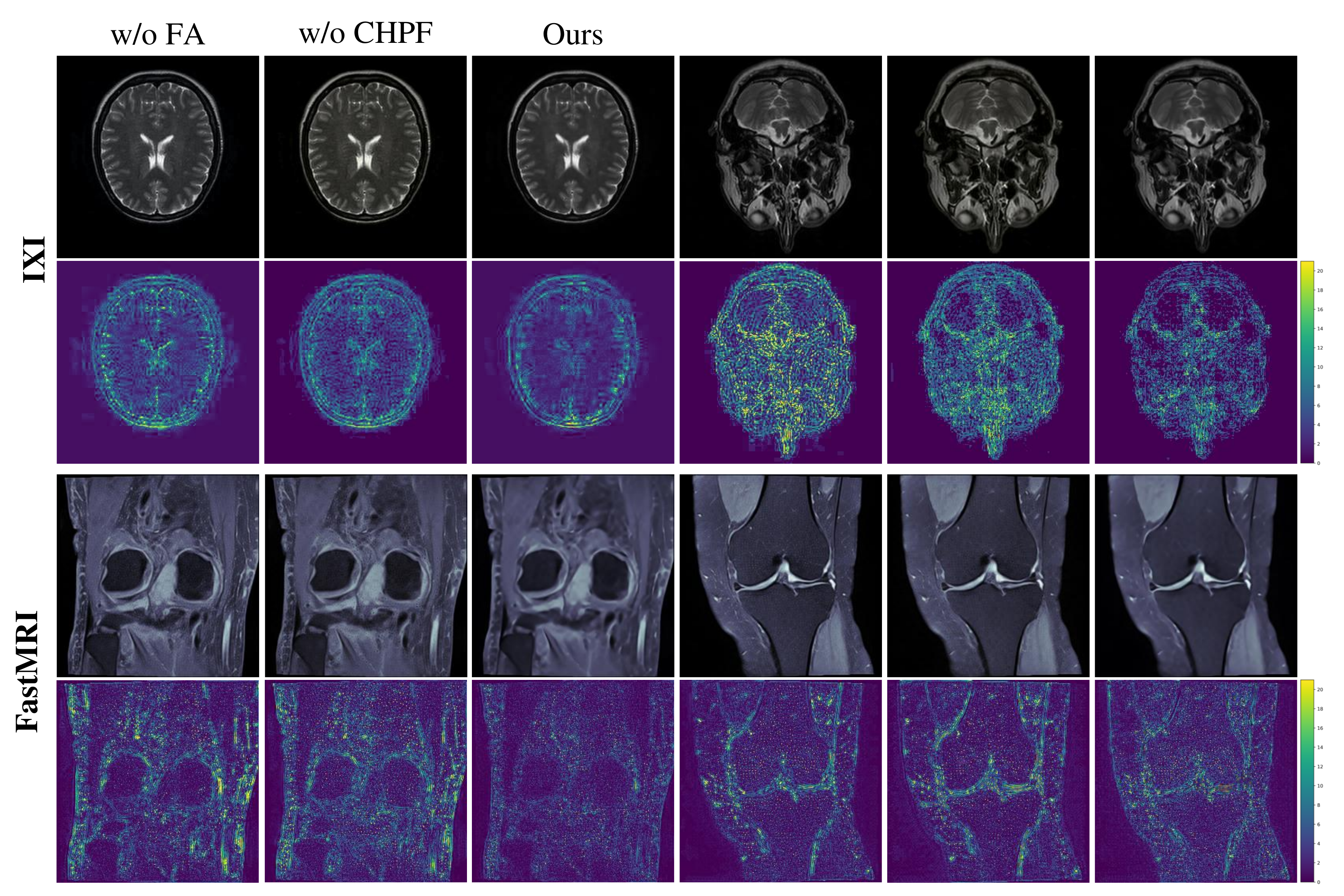}}
	\caption{\label{fig:wo}  Visual comparison of the output and residual maps from different models under 4$\times$ enlargement on the IXI and FastMRI dataset, respectively. }
\end{figure*}

\subsection{Ablation Study}
To find the optimal network structure, we established a study of $3$ sets of ablation experiments. In the first study, the effect of the backbone on the model is mainly compared. The second ablation experiment verifies the effectiveness of the proposed modules of this algorithm. The third ablation experiment evaluates the effect of different loss functions on the image inpainting effect.

\subsubsection{Shallow Feature Extractor Backbone Ablation}
\begingroup
\setlength{\tabcolsep}{6pt} 
\renewcommand{\arraystretch}{1.5} 
\begin{table}[htbp]
	\centering
	\caption{\label{tab:backbone}  Ablation Study of Shallow Feature Extractor Backbone}
	\begin{tabu}{ccccc}
		\tabucline[1.5pt]{-}
		{Backbone} & {\# Params$\left(M\right)$} & {\# FLOPs$\left(G\right)$} & PSNR & SSIM \\
		\hline \hline
		resnet18  & 12.87 & 32.04 & 38.55 & 0.959 \\
		resnet34  & 22.51 & 68.09 & 38.70 & 0.965 \\
		VGG19     & 12.34 & 22.06 & 38.52 & 0.974  \\
            VIT-B     & 86.34  &55.40  & 38.74 & 0.972  \\
            VIT-L     & 307.14 &190.70 & 38.75 & 0.975  \\
            Swin-B    & 88.01  &37.10  &38.71  & 0.977 \\
		\tabucline[1.5pt]{-}
	\end{tabu}
\end{table}
To determine the influence of different backbone architectures on PSNR and SSIM scores, we selected three widely used feature extractor networks and conducted three experiments on the IXI dataset.
We summarized the number of parameters, FLOPs, PSNR, and SSIM scores in \textcolor{cyan}{Tab.~\ref{tab:backbone}}. The vit-large feature extraction module achieves the best results because of its huge FLOPs and parameters. However, the improvement that vit can bring is limited, mainly because of the limited amount of the dataset. What’s more,~\cite{hassani2022neighborhood},~\cite{yoo2022rich} has proved the convolution structures are beneficial to cross-attention mechanisms. To sum up, considering that the promotion brought by vit is not big, we finally choose VGG structure as the backbone. To make the number of feature channels of each output layer of each model the
same, we modified the feature map channels of the network based on the vit source code. So the number of parameters and FLOPs will be inconsistent with those officially provided by vit.
In addition, the parameters in the feature extractor layer of each model are counted, and the FLOPs are quantified by processing a 256$\times$256$\times$3 shaped image.

\subsubsection{Effectiveness of Proposed Modules }
In the second set of ablation experiments, we control a single variable to evaluate the effectiveness of the proposed module, including 1) The Multi-Multi Pyramid Alignment module (M-A); 2) The Single-Multi Pyramid Alignment module (S-A); 3) The Cross-Hierarchical Progressive Fusion module (CHPF); 4) Flexible Alignment module (FA) combines the S-A and M-A modules. The quantitative results are shown in Tab.~\ref{tab:wo}, and the qualitative visual effects are shown in \textcolor{cyan}{Fig.  \ref{fig:wo}} .

In the W/O S-A experiment, we removed the S-A module so that the Flexible Alignment (FA) module retained only the M-A module. Compared with the experimental results of the complete structure, both PSNR and SSIM indexes are reduced. It is mainly because when the feature map of the reference image and low-resolution image have different scales in the receptive field, the module of S-A can adaptively balance it to ensure complete feature alignment. In contrast, W/O M-A removes the M-A module and only retains the S-A module in the FA module. The S-A module ensures that the patch foreground is consistent with the patch size to avoid excessive background blending. However, this group of experimental indicators declined more seriously, indicating that M-A had a better effect on the feature alignment than S-A. In the w/o FA experiment, the Flexible Alignment module, which includes both M-A and S-A alignment modules, was replaced by cross-attention. Owing to the limitation of the receptive field, cross-attention exhibits the worst performance in the alignment model, which also indicates the common effectiveness of S-A and M-A. The w/o CHPF model uses a decoder to directly restore the aligned features and ignores the fusion of multi-hierarchical feature maps, which also has a poor experimental effect on the IXI testset.

\begingroup
\setlength{\tabcolsep}{10pt} 
\renewcommand{\arraystretch}{1.5} 
\begin{table}[!htbp]
	\centering
	\caption{\label{tab:wo} Quantitive comparision of proposed modules on the ixi dataset}
	\begin{tabu}{c|ccc|cc} 
		\tabucline[1.5pt]{-}
		& S-A & M-A & CHPF & PSNR & SSIM  \\ 
		\hline \hline
		w/o S-A &               &  \checkmark  & \checkmark & 38.12 &	0.961   \\
		w/o M-A &\checkmark	    &              & \checkmark & 37.93 &  0.953  \\
		w/o CHPF  &	\checkmark             &  \checkmark             & & 35.14  &	0.907  \\
		w/o FA  & 	&    & \checkmark            & 35.06 &  0.898  \\
		complete   &\checkmark   & \checkmark   & \checkmark  & 38.52&	0.974   \\
		\tabucline[1.5pt]{-}
	\end{tabu}
\end{table}

\subsubsection{Loss Ablation}
The third set of ablation experiments is to verify the impact of different loss functions on the quality of restored images. As seen in Tab.~\ref{tab:loss}, in the absence of L1 loss, the effect of the model falls off dramatically. This is mainly because in the Super-Resolution process, the pixel-by-pixel loss can largely ensure the stability of the image structure to be repaired and the authenticity of the details, which also demonstrates the importance of L1-loss. In addition, when lacking perceptual loss,the performance of image restoration is also reduced. It is because perceptual loss supervises the reconstructed image at the semantic level so that the network can pay more attention to the accuracy of image semantic information and macroscopically control the trend of image restoration.

Furthermore, in the absence of Fourier loss, it is found that the SSIM score decreases more than PSNR. It is primarily attributed to the that it is easy to operate the high-frequency and low-frequency information of the image separately in the frequency domain, which significantly improves the image's contrast,so it has a more significant impact on the SSIM score. Note that the outcomes are acquired using VGG19 as the feature extractor backbone.

\begingroup
\setlength{\tabcolsep}{10pt} 
\renewcommand{\arraystretch}{1.5} 
\begin{table}[!htbp]
	\centering
	\caption{\label{tab:loss} Ablation Study of  Loss Function}
	\begin{tabu}{ccc|cc} 
		\tabucline[1.5pt]{-}
		$\mathcal{L}_1$ & $ \mathcal{L}_{ssim}$ & $\mathcal{L}_{FR}$ & PSNR & SSIM  \\ 
		\hline \hline
		& \checkmark &  \checkmark&  36.84 &	0.966   \\
		\checkmark	&  & \checkmark & 38.11 &	0.925   \\
		\checkmark	& \checkmark &  & 37.07 &	0.967  \\
		\checkmark	& \checkmark & \checkmark &  38.52 & 	0.974  \\
		\tabucline[1.5pt]{-}
	\end{tabu}
\end{table}

\begingroup
\setlength{\tabcolsep}{10pt} 
\renewcommand{\arraystretch}{1.5} 
\begin{table}[!htbp]
	\centering
	\caption{\label{tab:noise} Noise and Artifact Robustness Study}
	\begin{tabu}{ccc|cc} 
		\tabucline[1.5pt]{-}
		grade & motion & RF & PSNR & SSIM  \\ 
		\hline \hline
		light & \checkmark &  &  37.18 &	0.968   \\
		strong	&\checkmark  & & 29.81 &	0.902   \\
		light	& &  \checkmark& 37.05 &	0.967  \\
		strong	&  & \checkmark &  30.57 & 	0.911  \\
		\tabucline[1.5pt]{-}
	\end{tabu}
\end{table}

\subsubsection{Noise Robustness}
Here, we present the performance of the model against common instrumental image interference noise and artifacts present in MRI images. The common interference noise and artifacts in MRI images are mainly caused by motion and radio frequency (RF) interference. We used motion blur function and local periodic interference signal to simulate noise and artifacts in input images. In the presence of slight noise, as shown in Tab.~\ref{tab:noise}, the model has a good performance on noise robustness. When the noise is enhanced, the performance of the model shows a certain decline. Subsequently, we will improve the performance of S-A and M-A modules in the noise conditions.

\section{Conclusion}
In this paper, we have proposed the Flexible Alignment Super-Resolution network for Multi-Contrast MRI images (FASR-Net), which utilizes a related reference image to obtain a more authentic output.

Specifically, the Flexible Alignment (FA) Module was proposed to align the semantic features extracted from Ref and LR images by a pre-trained VGG19 network. FA module can be subdivided into the Multi-Multi Pyramid Alignment (M-A) module and the Single-Multi Pyramid Alignment 
(S-A) module. The M-A module is responsible for aligning input images with the same scale, and S-A deals with the images with different scale. In addition, the Cross-Hierarchical Progressive Fusion (CHPF) module was proposed to progressively fuse the aligned features generated by FA module. At last, extensive experiments on the IXI and the FastMRI datasets proved our method could retain more textural details by comparing with the counterparts obtained by the existing methods in the visual comparison figures.
\bibliographystyle{IEEEtran}
\bibliography{egbib}
\end{document}